\documentclass[11pt]{article}
\pdfoutput=1 
\usepackage{jinstpub} 
\usepackage{subfigure}

\title{Helical FOFO Snake for Initial Six-Dimensional Cooling of Muons}

\author[a]{Yuri Alexahin}

\affiliation[a]{Fermi National Accelerator Laboratory,\\Batavia, IL 60510, USA}

\emailAdd{alexahin@fnal.gov}

\abstract{The helical FOFO snake six-dimensional muon ionization cooling channel design is presented which incorporates wedge absorbers in such a way that simultaneous cooling of both signs of muons is possible.}

\begin{document}
\maketitle
\flushbottom

\section{Introduction}

The major difficulty with ionization cooling of muons is anti-damping of longitudinal oscillations. This is due to the decrease of ionization loss with momentum in the range most suitable for cooling, \mbox{2--300}~MeV/$c$. A number of schemes have been proposed to resolve this difficulty and provide six-dimensional (6D) cooling by forcing muons with higher momentum to take a longer path in the absorber so that they lose more energy. This can be realized by creating dispersion in particle positions at a wedge absorber (without significant overall path lengthening) or by creating sufficiently large path lengthening with momentum and using a homogeneous absorber.

The early versions of the so-called FOFO snake~\cite{fofo} used a third possibility: locally large path lengthening in slab absorbers due to a large slope of the dispersion function. This allows the FOFO snake to cool muons of both signs simultaneously.

Recent work has shown that a ``helical'' (HFOFO) snake can incorporate wedge absorbers in such a way that simultaneous cooling of $\mu^{+}$ and $\mu^{-}$ is still possible. This enables smaller ``snake'' amplitude and improved transmission.

\section{Basic Principles}

The helical FOFO snake is based on the following principles: alternating solenoid focusing, periodic rotating dipole field and resonant dispersion generation~\cite{fofo}.

The focusing magnetic field is created by a sequence of solenoids with alternating polarity with gaps between them (the name FOFO reflects the fact that solenoid focusing does not depend on polarity since it is quadratic in the magnetic field). Emittances of the two transverse normal modes (in the axisymmetric field they are the cyclotron (Larmor) and drift modes~\cite{burov}) are swapped with each change of polarity so that both modes are cooled.

The transverse magnetic field component necessary for dispersion generation can be created by a periodic inclination of the solenoids. The idea of the helical FOFO snake is to make a rotating dipole field by inclining solenoids in rotating planes $x\cos(\phi_k) + y\sin(\phi_k)\!=\!0$, $\phi_k\!=\!\pi(1-2/N_s)(k\!+\!1)$, $k=1,2,\ldots,\!N_s$, where $N_s$ is the number of solenoids per period. 

If $N_s=2(2j+1)$, where $j$ is an integer, then a $\mu^{-}$ in solenoid $k$ sees exactly the same forces as a $\mu^{+}$ in solenoid $k\!+\!N_s/2$, since these solenoids have the same inclination but opposite polarity. As a result, $\mu^{-}$ and $\mu^{+}$ orbits have exactly the same form with a longitudinal shift by a half-period ($N_s/2$ solenoids), but are not mirror-symmetric as one might expect.

A large dispersion can be generated if the transverse tune $Q_{\perp}$ is close to a resonant value. To obtain a positive momentum compaction favorable for longitudinal cooling, it must be above the resonant value $Q_{\perp} > n\!+\!Q_s$, where $Q_s$ is the longitudinal mode tune. Despite the closeness to a resonance, the momentum acceptance of such a channel can be sufficiently large owing to higher order chromatic effects.

\section{Lattice Description}
Here we present a version of the helical FOFO snake filled with high-pressure gaseous hydrogen (GH$_2$). The high-pressure hydrogen acts as the cooling channel absorber and also allows higher RF gradient. Following the HCC design~\cite{hcc}, we assume its density to be equal to 20\% that of liquid hydrogen and take for the peak RF gradient $E_{max}\!=\!25$~MV/m at $f_{RF}\!=\!325$~MHz. It is assumed that GH$_2$ will provide sufficient cooling of the cavities' windows so that they can be quite thin (0.12 mm of Be at the beginning of the channel).

\begin{figure}[htbp]
\centering
 \includegraphics[width=\textwidth]{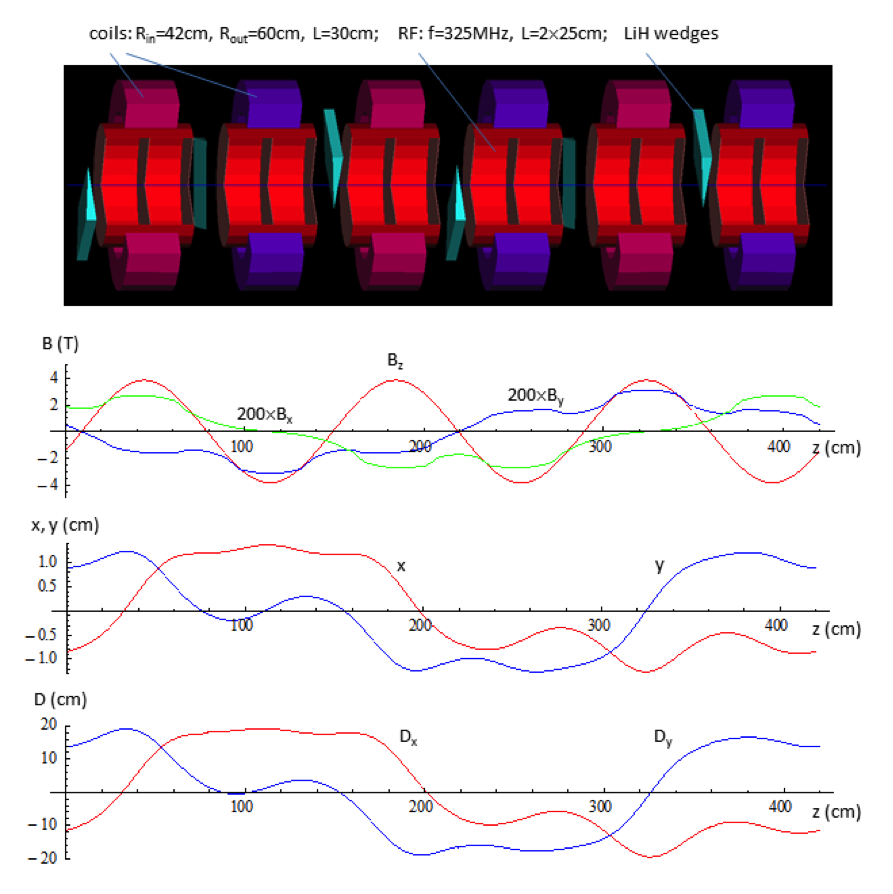}
 \caption{Layout of one period of the HFOFO lattice (top), magnetic field for muon momentum 230 MeV/c (second from top), $\mu^{+}$ equilibrium orbit and dispersion (bottom).}
 \label{fig:hfofo}
\end{figure}

One period of the channel is schematically shown in Figure~\ref{fig:hfofo} (top). Its length is $L_{\text{period}}=4.2$~m. There are $N_s = 6$ solenoids per period, each inclined by 2.5~mrad 
about axes that rotate about the  channel axis by $\phi_k=4\pi/3, 0, 2\pi/3, 4\pi/3, 0, 2\pi/3$, where $\phi = 0$ corresponds to inclination 
about the horizontal axis.

The relatively small radius of the RF cavities at $f_{RF}\!=\!325$~MHz made it possible---in contrast to the earlier versions of the FOFO snake~\cite{fofo}---to place the cavities inside the solenoids and to free up space for LiH wedge absorbers (wedge angle 0.17 rad at the beginning of the channel) between the solenoids at the minima of the betatron functions. To ensure equal longitudinal cooling for $\mu^{-}$ and $\mu^{+}$, the wedges with numbers $k$ and $k\!+\!N_s/2$, $k=1,2,3$ are grouped in pairs of the same orientation. This orientation was then optimized to achieve maximum longitudinal cooling.

\section{Properties of Periodic Channel}
The lower two plots in Fig.~\ref{fig:hfofo} show the $\mu^{+}$ equilibrium orbit and dispersion found for a momentum of 230~MeV/$c$. The normal mode tunes and normalized equilibrium emittances are given in Table~\ref{tab:hfofo_tunes}.

\begin{table}[htbp]
\caption{\label{tab:hfofo_tunes} Normal mode tunes and normalized equilibrium emittances.}
\centering
\begin{tabular}{|cccc|}
\hline
Parameter & Mode I & Mode II & Mode III \\
\hline
Tune & $1.2271 + 0.0100i$ & $1.2375 + 0.0036i$ & $0.1886 + 0.0049i$ \\
Emittance (mm) & 2.28 & 6.13 & 1.93 \\
\hline
\end{tabular}
\end{table}

The tunes were computed from eigenvalues of a one-period transfer matrix; their imaginary part describes oscillation damping due to the regular part of the ionization loss. Without stochastic effects, the emittances would have been damped as
\[
  \frac{d}{ds} \ln \varepsilon_j = -2 \times \frac{2\pi}{L_{\text{period}}} \text{Im} Q_j.
\]

\begin{figure}[htbp]
\centering
 \includegraphics[width=0.8\textwidth]{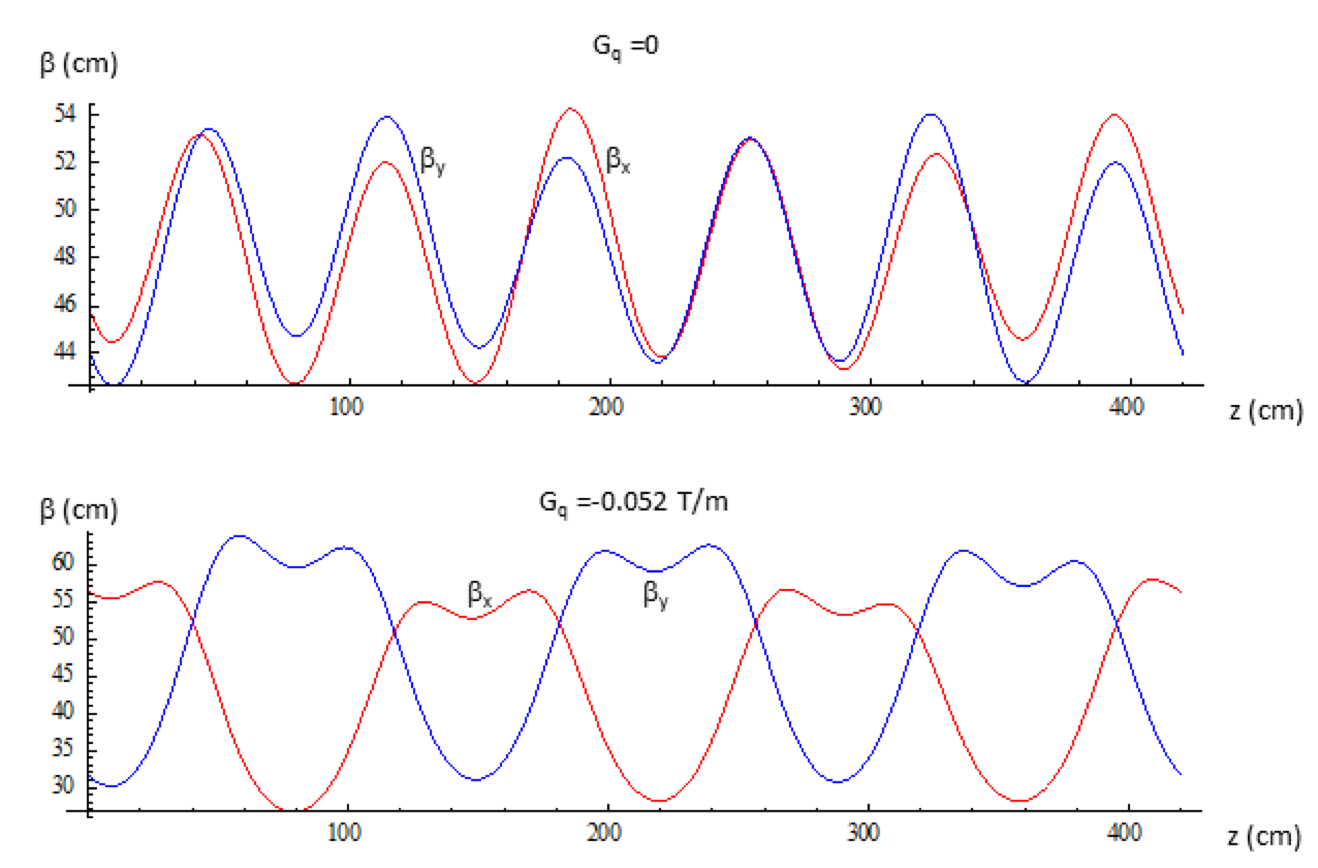}
 \caption{$\mu^{+}$ transverse $\beta$-functions with no (top) and with constant quadrupole field of $G_q=-0.052$\,T/m (bottom).}
 \label{fig:hfofo_beta_quad}
\end{figure}

There is a large difference between the cooling rates and equilibrium emittances of the transverse normal modes (I and II) due to the axial symmetry breaking by the dipole field component. They can be equalized with the help of a unipolar quadrupole field (not necessarily of a constant gradient). Such a field works for both $\mu^{-}$ and $\mu^{+}$ despite breaking the translational symmetry between the two beams. However, a strong beta-beat is excited (see Fig.~\ref{fig:hfofo_beta_quad}) which slightly increases the 4D emittance. In the version of the channel presented here the quadrupole field is turned off.

\begin{figure}[htbp]
\centering
 \includegraphics[width=0.8\textwidth]{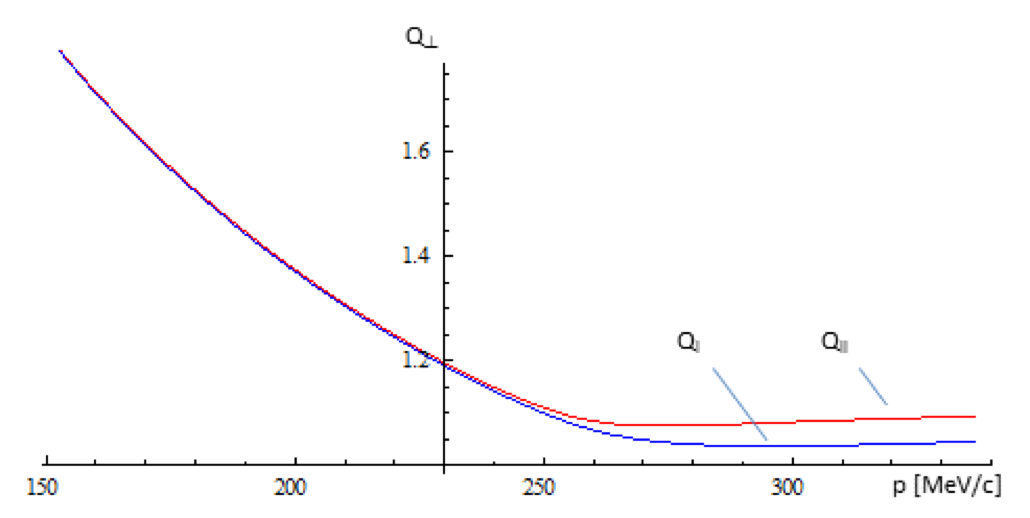}
 \caption{Betatron tunes as function of muon momentum.}
 \label{fig:hfofo_tunes}
\end{figure}

To estimate the momentum acceptance, we can assume the longitudinal motion to be slow and calculate betatron tunes for a constant momentum $p$ (Figure~\ref{fig:hfofo_tunes}). Surprisingly, the transverse tunes are ``repelled'' from the integer resonance making what we may call the static acceptance very large (it significantly exceeds the shown range). However, there is a limitation due to a change in the sign of the slippage factor~\cite{fofo}. For the parameters under consideration this sets the upper limit of what can be called the dynamic acceptance at $\sim$333~MeV/$c$ while its
lower limit (defined by the parametric resonance, $Q_{\textnormal{perp}}=1.5$) is at $\sim$180~MeV/$c$.

\section{Initial 6D Cooling Simulations}
The HFOFO snake can be used for cooling $\mu^{-}$ and $\mu^{+}$ beams produced in the front end~\cite{325}. The average momentum of the beam core from the front end is rather high, $\sim$250~MeV/$c$, so in order to pull it farther from the upper limit and reduce losses, the design momentum was gradually lowered along the channel to $\sim$200~MeV/$c$. This was achieved by lowering the current in the solenoids and adjusting the RF phase and LiH absorber wedge angle while keeping the solenoid geometry and RF gradient constant. The total length of the channel including matching sections at both ends is 131~m.

\begin{figure}[htbp]
\centering
 \includegraphics[width=\textwidth]{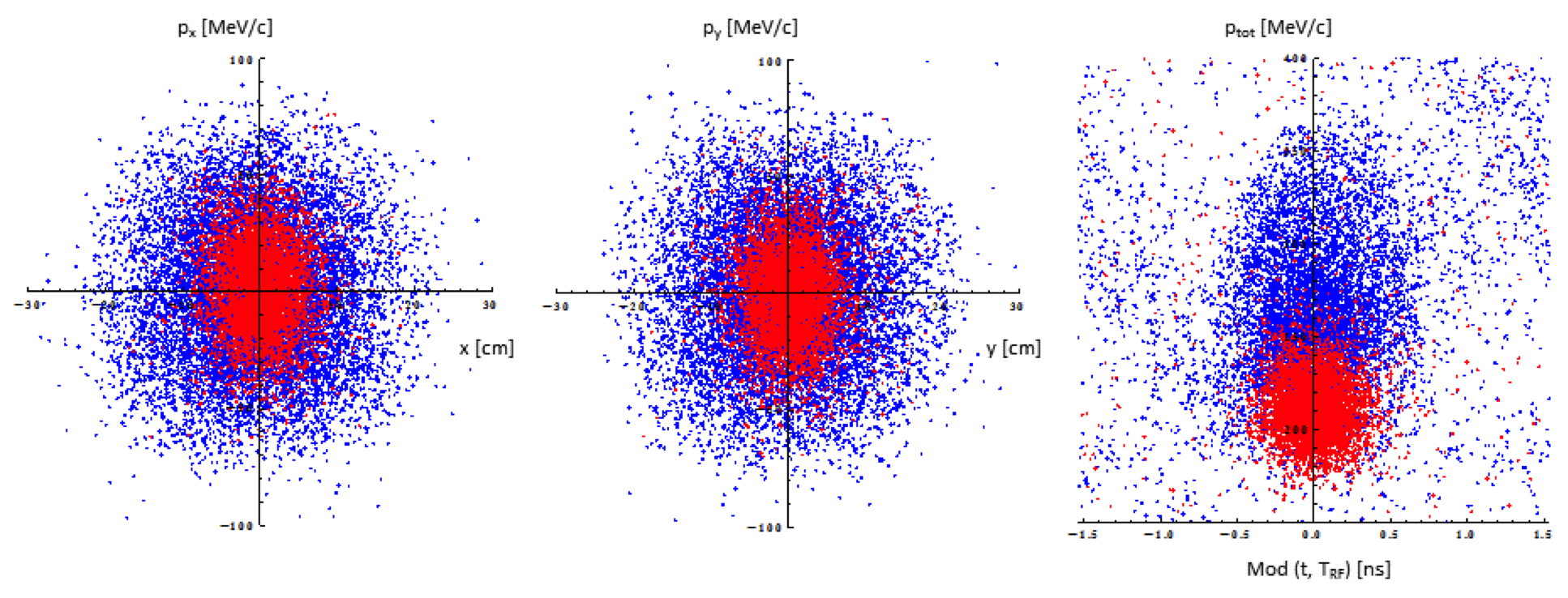}
 \caption{Initial (blue) and final (red) $\mu^{+}$ beam phase space distribution. All bunches were projected onto the same RF bucket in the right plot.}
 \label{fig:hfofo_cooling}
\end{figure}

\begin{figure}[htbp]
\centering
 \includegraphics[width=0.8\textwidth]{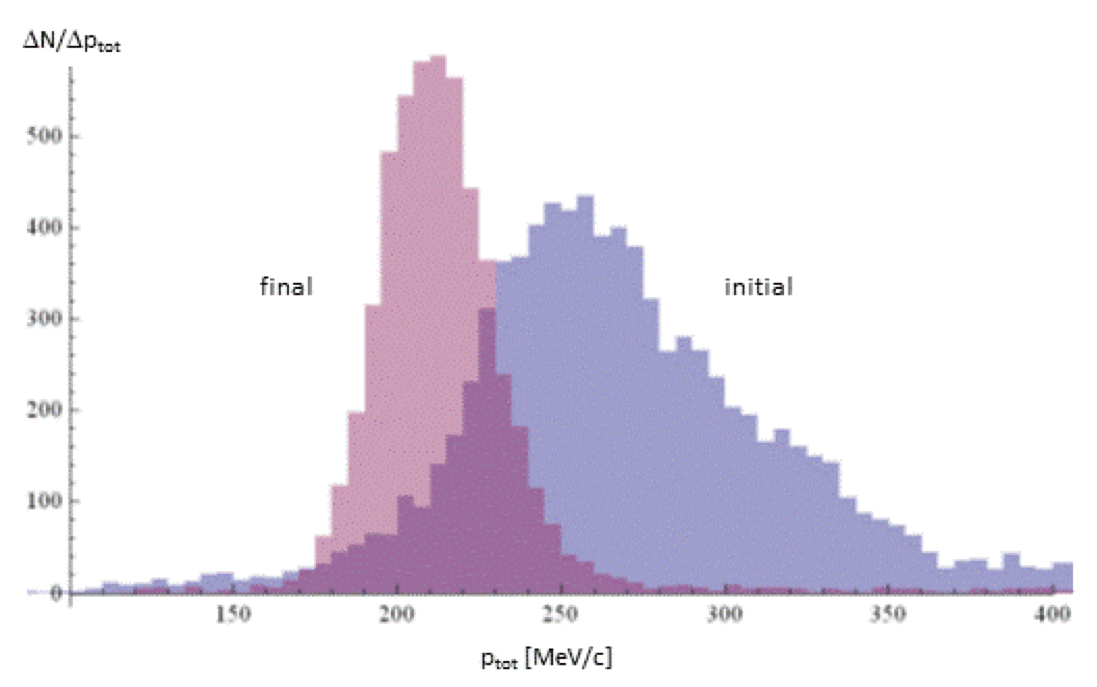}
 \caption{Initial (blue) and final (pink) $\mu^{+}$ beam distribution in total mechanical momentum}
 \label{fig:hfofo_momentum}
\end{figure}

Figures~\ref{fig:hfofo_cooling} and \ref{fig:hfofo_momentum} show the distributions of the initial $\mu^{+}$ beam and the cooled beam in the exit solenoid (with the same 2 T magnetic field as in the front end solenoid) obtained by tracking with G4beamline~\cite{g4beamline}. No cuts were applied. Distributions of the $\mu^{-}$ beam look similar.

\begin{figure}[htbp]
\centering
 \includegraphics[width=\textwidth]{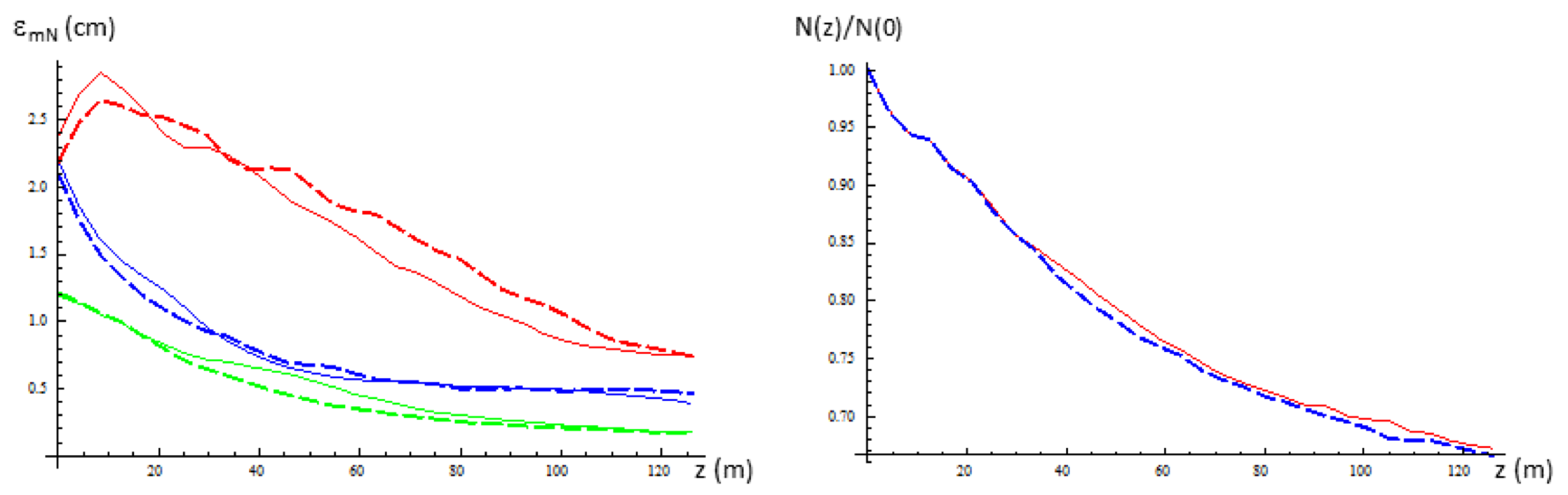}
 \caption{Normal mode emittances (left) and beam core intensity (right) for $\mu^{+}$ (solid lines) and $\mu^{-}$ (dashed lines) along the HFOFO channel.}
 \label{fig:hfofo_emittance_evolution}
\end{figure}

Computation of the beam emittance presents a challenging problem due to long non-Gaussian tails in the distribution. To obtain unambiguous results, we use a multi-dimensional Gaussian fit~\cite{eigenemittances}, which automatically suppresses the halo contribution. Computed in this way, normalized emittances and beam core intensities are shown in Fig.~\ref{fig:hfofo_emittance_evolution} for the case when no quadrupole field was added which leaves the equilibrium emittances unequalized. The initial 6D emittances 5.6 cm$^3$ ($\mu^{-}$) and 6.2 cm$^3$ ($\mu^{+}$) were reduced to 0.051 cm$^3$ for both beams.

\section{Summary \& Outlook}
 
It is possible to achieve 6D ionization cooling of both $\mu^{+}$ and $\mu^{-}$ simultaneously 
in a straight GH$_2$-filled channel with a helical dipole field and wedge absorbers (HFOFO snake).
The present design based on the 325\,MHz RF has sufficient transverse and longitudinal acceptance
to serve as the initial stage of the 6D cooling channel of the muon collider.

The initial/final 6D emittance ratio achieved in a 124\,m long channel is 112.8 (36.4 in the
transverse 4D emittance and 3.1 in the longitudinal emittance) with a transmission of ~65\%.

\section{Comments}

In the present HFOFO snake design, the solenoid inclination was chosen as the source of the transverse dipole field. It would be more practical, however, to use separate independently powered coils for this purpose.

The 6D emittance can be further reduced by a factor of $\sim$5 by tapering down the length of all elements and increasing the solenoid current to keep the transverse tunes intact.
 
A more significant emittance reduction can be achieved with higher betatron phase advance per focusing unit (solenoid plus adjacent gap) than the $\phi=74^\circ$ in the present design. This will reduce the beta-function at the minima reached between solenoids for $\phi<180^\circ$ and at the solenoid centers for $\phi>180^\circ$. To fully benefit from such a reduction, the absorbers should be localized at the beta-function minima making it necessary to use vacuum RF or, after merge when there is only one bunch per beam, pulsed radial lines~\cite{caporaso}.

\section*{Acknowledgments}
The author is grateful to D. Kaplan and P. Snopok (IIT, CHicago) for helpful discussions. This work was supported by Fermilab under contract No.\ DE-AC02-07CH11359 with the U.S. Department of Energy.

\end{document}